 \documentstyle[preprint,eqsecnum,aps]{revtex}

\begin{document}
\draft
\preprint{hep-th/9410244}
\title{A Solution of the Maxwell-Dirac Equations in 3+1 Dimensions
}
\author{A. Garrett Lisi}
\address{
Department of Physics, University of California San Diego, La Jolla, CA
92093-0319
}
\date{October 31, 1994}
\maketitle
\begin{abstract}
We investigate a class of localized, stationary, particular numerical
solutions to the Maxwell-Dirac system of classical nonlinear field
equations.  The solutions are discrete energy eigenstates bound
predominantly by the self-produced electric field.
\end{abstract}

\narrowtext

\section{Introduction}

There are many examples of classical solitary wave solutions to nonlinear
field theory equations.  Some of these are useful in quantum field theory
as a stationary point in the action functional about which one quantizes
the field \cite{raj}.  Although these solutions are often relegated to
model equations in fewer than three space dimensions \cite{das}, we
consider the Maxwell-Dirac system of equations in 3+1 dimensional
space-time, a nonlinear system of PDE's involving twelve real functions of
four variables.  The general solution of this system is certainly well
beyond our grasp; however, we will obtain a class of particular solutions
by making simplifying assumptions and utilizing numerical methods.

Before we embark on a search for a localized solution it would be wise to
consider on what grounds such a solution is plausible.  It is well known
that when one considers the Dirac equation with an external ``repulsive''
potential the possibility arises to obtain bound state solutions
\cite{sak}.  This potential produces bound states with discrete energies
that rise from the continuum of free negative energy states in the same
way that an ``attractive'' potential produces lower energy states from
those of positive energy.  For a great enough ``repulsive'' potential we
may even obtain bound states of positive energy.  This interesting
phenomenon goes under the name of ``Klein's paradox'' and provides our
motivation.  In our case the ``repulsive'' potential is provided by the
charge feeling its own electric field.

\section{Field Equations}

The Maxwell-Dirac equations are obtained from the Lagrangian density
\begin{equation}
{\it L} = i  \bar{\Psi} {\gamma}^{\mu} {\partial}_{\mu} {\Psi}
-  \bar{\Psi} {\Psi}
- q \bar{\Psi} {\gamma}^{\mu} {\Psi} {A}_{\mu}
- {1 \over 4} {F}^{\mu \nu} {F}_{\mu \nu}
, \label{mdl}
\end{equation}
in which the c-number fields are the Dirac spinor $\Psi$, which can be
considered a four component single-column matrix, and we have
${A}^{\mu}=(\Phi,\vec{A})$, $\bar{\Psi} \equiv {\Psi}^{\dagger}
{\gamma}^{0}$, ${F}^{\mu \nu}\equiv{\partial}^{\mu}{A}^{\nu} -
{\partial}^{\nu}{A}^{\mu}$, ${\partial}_{\mu}\equiv{{\partial} \over
{\partial {x}^{\mu}}}$, ${x}^{\mu}=(t,\vec{x})$, and ${\gamma}^{\mu}$ are
the 4$\times$4 matrices
\begin{eqnarray*}
{\gamma}^{0} \equiv \pmatrix{
I & 0 \cr
0 & -I \cr
},
\vec{\gamma} \equiv  \pmatrix{
0 & \vec{\sigma} \cr
-\vec{\sigma} & 0 \cr
},
\end{eqnarray*}
in which $I$ is the 2$\times$2 identity matrix and $\vec{\sigma}$
represents the Pauli matrices
\begin{eqnarray*}
{\sigma}_{x} \equiv \pmatrix{
0 & 1 \cr
1 & 0 \cr
},
{\sigma}_{y} \equiv \pmatrix{
0 & -i \cr
i & 0 \cr
},
{\sigma}_{z} \equiv \pmatrix{
1 & 0 \cr
0 & -1 \cr
}. \end{eqnarray*}
The Euler-Lagrange equations applied to Eq.\ (\ref{mdl}) give the
Maxwell-Dirac equations
\begin{equation}
{\gamma}^{\mu} (i {\partial}_{\mu}  - q  {A}_{\mu}) {\Psi}
-  \Psi
=  0, \label{de}
\end{equation}
\begin{equation}
{\partial}_{\nu} {F}^{\mu \nu}
= q \bar{\Psi} {\gamma}^{\mu} {\Psi}. \label{me}
\end{equation}
Throughout we use natural units, in which we have rescaled length, mass,
and time so that $\hbar=m=c=1$.

We now begin making assumptions about the solution we wish to look for.
We require that ${A}^{\mu}$ satisfy the Lorentz condition
${\partial}_{\mu} {A}^{\mu} = 0$ and that $\Psi$ is an energy eigenstate,
$\Psi=\psi(\vec{x}) {e}^{-i E t}$.  Equation\  (\ref{me}) now reduces to
Poisson's equation
\begin{equation}
{\nabla}^{2} {A}^{\mu} = - {j}^{\mu} , \label{pe}
\end{equation}
where ${j}^{\mu}$ is the 4-current
\begin{equation}
{j}^{\mu} = q \bar{\psi} {\gamma}^{\mu} {\psi}, \label{j4}
\end{equation}
and Eq.\ (\ref{de}) can be written as a Hamiltonian eigenvalue equation
\begin{equation}
E {\psi} = H {\psi}  = [{\gamma}^{0} \vec{\gamma} \cdot (-i \vec{\nabla} -
q \vec{A} ) + {\gamma}^{0}  + q \Phi]{\psi} . \label{vde}
\end{equation}

We would now like to assume spherical symmetry for our wave function;
however, we find that the resulting vector potential $\vec{A}$ has an
angular dependence that destroys the symmetry.  This gives us two options:
we may throw out the magnetic term in the hope that its contribution will
be small and solve the resulting one dimensional, spherically symmetric
problem, or we may attack the non-spherical problem.  We proceed with the
former and save the latter for Sec.\ \ref{sec5}.  By removing $\vec{A}$ we
reduce the Maxwell-Dirac system (\ref{mdl}) to a massless scalar-Dirac
system
\begin{equation}
{\it L} = i \bar{\Psi} {\gamma}^{\mu} {\partial}_{\mu} {\Psi}
-  \bar{\Psi} {\Psi}
- q \bar{\Psi} {\gamma}^{0} {\Psi} \Phi
+ {1 \over 2} ({\partial}_{\mu} \Phi)({\partial}^{\mu} \Phi)
. \label{sdl}
\end{equation}
Note that, with $\gamma^0$ in the coupling term, our massless scalar-Dirac
system (\ref{sdl}) is not Lorentz invariant and is therefore only of use
in approximating the Maxwell-Dirac system.

The necessity for a spherical charge distribution restricts our
wavefunction to four possible configurations corresponding to total
angular momentum up or down, $ m_z=\pm {1 \over 2} $, and the quantum
number $ \kappa=\pm1 $,
\begin{eqnarray}
\matrix{
\kappa = -1 \cr
m_z = {1 \over 2} \cr}
\  \rm{for} \  \psi &=& \left (\matrix{
g(r)\cr
0\cr
-i f(r) \cos{\theta} \cr
-i f(r) {e}^{i \phi} \sin{\theta} \cr
}\right ), \nonumber \\
\matrix{
\kappa = 1 \cr
m_z = {1 \over 2} \cr}
\  \rm{for} \  \psi &=& \left (\matrix{
g(r) \cos{\theta} \cr
g(r) {e}^{i \phi} \sin{\theta} \cr
-i f(r)\cr
0\cr
}\right ),\nonumber  \\ \matrix{
\kappa = -1 \cr
m_z =- {1 \over 2} \cr}
\  \rm{for} \  \psi &=& \left (\matrix{
0\cr
g(r)\cr
-i f(r) {e}^{-i \phi} \sin{\theta} \cr
i f(r) \cos{\theta} \cr
}\right ), \nonumber  \\
\matrix{
\kappa = 1 \cr
m_z = -{1 \over 2} \cr}
\  \rm{for} \  \psi &=& \left (\matrix{
g(r) {e}^{-i \phi} \sin{\theta} \cr
-g(r) \cos{\theta} \cr
0\cr
-i f(r)\cr
}\right ), \label{swave}
\end{eqnarray}
in spherical coordinates $(r,\theta,\phi)$, where we have chosen to orient
the angular momentum along the z-axis.  Note that the two $\kappa=-1$
configurations have the same angular dependence as the Hydrogen ground
state.  Equations (\ref{pe}-\ref{vde}), neglecting $\vec{A}$, now reduce
to the radial equations
\begin{eqnarray}
E \left (\matrix{ g \cr f \cr }\right ) &=& \pmatrix{
q \Phi + 1 &  -{d \over {d r}}+\frac{\kappa-1}{r}  \cr
{d \over {d r}}+\frac{\kappa+1}{r}  & q \Phi - 1 \cr
}
\left (\matrix{ g \cr f \cr }\right ), \label{fgode} \\
\Phi^{\prime \prime} + {2 \over r} \Phi^\prime &=& - q ({f}^{2}+{g}^{2}).
\label{uode}
\end{eqnarray}
We symmetrize the Hamiltonian in Eq.\ (\ref{fgode}) by the similarity
transformation $F \equiv r f,G \equiv r g$, and further simplify Eqs.\
(\ref{fgode}) and (\ref{uode}) by identifying the fine structure constant
$\alpha \equiv {{q}^{2} \over {4 \pi }}$ and defining the potential $ V
\equiv q \Phi$, giving us
\begin{eqnarray}
E \left (\matrix{ G \cr F \cr }\right ) &=& \pmatrix{
V + 1 & -{d \over {d r}}+ \frac{\kappa}{r}  \cr
{d \over {d r}}+\frac{\kappa}{r}  & V - 1 \cr
}
\left (\matrix{ G \cr F \cr }\right ), \label{wode} \\
 (r V)^{\prime \prime} &=& - {{4 \pi \alpha} \over r} ({F}^{2}+{G}^{2}).
\label{ode}
\end{eqnarray}
We also restrict our wavefunction by imposing the normalization condition
\begin{equation}
1 = \int{{d}^{3}x {\Psi}^{\dagger}} \Psi = 4 \pi \int_{0}^{\infty}{dr
({F}^{2}+{G}^{2})}, \label{nrm}
\end{equation}
ensuring a total charge of $q$.  Note that $m_z$ does not appear in Eqs.\
(\ref{wode}-\ref{nrm}); hence the energy levels are independent of the
choice $m_z = \pm {1 \over 2}$.

\section{Solution}

We may readily solve our system of O.D.E.'s\ (\ref{wode}, \ref{ode}) by a
variety of means, including power series solution \cite{rose}, Pad\'{e}
series approximation \cite{frac}, and numerical methods.  We choose the
latter as it seems the most straightforward path to obtaining a solution.

We discretize Eqs.\ (\ref{wode}) and (\ref{ode}) using second order finite
differences to obtain a standard, linear, symmetric, matrix eigenvalue
problem for $(G(r),F(r))$ and $E$ coupled nonlinearly to a symmetric,
matrix inverse problem for $r V$.  We obtain solutions by solving each
linear problem independently and iterating to convergence from an initial
guess.  This technique is similar to Newton's method of solving the
nonlinear system but allows for better behavior in solving the eigenvalue
problem at the cost of slower convergence.  We use inverse iteration for
the eigenvalue problem together with the conjugate gradient method
\cite{nrc} for the matrix inverses.  Although more efficient methods are
available for the one dimensional problem \cite{sal}, we have chosen
methods which will work equally well for the case of two independent
variables discussed in Sec.\ \ref{sec5}.

Figure~\ref{fg} shows a solution to our system\ (\ref{wode}-\ref{nrm}) for
the choice of $\alpha = 2.7$.  We may use the numerical solution to
calculate expectation values, such as $\left\langle r \right\rangle$ and
the strength of the neglected magnetic interaction.  Table \ref{table1}
shows the results for various selections of the parameter $\alpha$, which
completely determines the set of solutions.  The energy $E$ as a function
of $\alpha, \kappa$, and the number of nodes in $g$, represented by $n$,
is shown in Fig.~\ref{e}.  We can see from Fig.~\ref{e} and directly from
Eq.\ (\ref{wode}) that localized solutions are only possible for $-1<E<1$.
 Note that positive energy solutions exist for $\alpha>2.4$ and that there
is a spectrum of large $n$ states of energies near negative one for any
choice of $\alpha$.

\section{Magnetic Interaction Perturbation}
\label{sec4}

For the solution to our scalar-Dirac equations to be a viable approximate
solution to the Maxwell-Dirac equations, we must establish that the
inclusion of the magnetic interaction does not significantly alter the
solution.  We do this properly in Sec.\ \ref{sec5} by modeling the larger
system\ (\ref{pe}-\ref{vde}), including the angular dependence; however,
we will use perturbation theory with our spherically symmetric solution to
determine the approximate energy shift ${\Delta E}_{M}$ due to the
magnetic interaction and find good agreement with the full Maxwell-Dirac
case.

The magnetic interaction term in the Hamiltonian density is
\begin{equation}
{\it H}_M = -  \vec{j}\cdot \vec{A}. \label{hm}
\end{equation}
We use our solution form (\ref{swave}) in Eq.\ (\ref{j4}) to get the
current
\begin{equation}
{\vec j} =  4 q \kappa m_z f(r) g(r) \sin{\theta} \hat{\phi}. \label{jv}
\end{equation}
{}From this we obtain the vector potential ${\vec A} =  A(r)  \sin{\theta}
\hat{\phi}$, in which $A$ is calculated via the Green's function integral
\begin{equation}
A(r) = {1 \over 3} \int_{0}^{\infty}
{d {r}_{1}
{r}_{1}^{2}
{{{r}_{<}} \over {{r}_{>}}^{2}}
j(r_1)
}, \label{gfe}
\end{equation}
where $r_<$ ($r_>$) represents the lesser (greater) of $r$ and $r_1$.  We
may now calculate the approximate magnetic energy shift to first order via
\begin{eqnarray}
&&{\Delta E}_M \simeq \int{d^3 x  {\it H}_M} \\ \nonumber
&&= -{128 \over 9} \pi^2 \alpha
\int_{0}^{\infty}{dr F(r) G(r) { \int_{0}^{\infty}
{d {r}_{1}
{{{r}_{<}} \over {{r}_{>}}^{2}}
F(r_1) G(r_1)
}}}. \label{em}
\end{eqnarray}
The values obtained for ${\Delta E}_M$ are seen in Fig.~\ref{e} and Table
\ref{table1} and found to be small compared to the binding energy.

We may also use our current (\ref{jv}) to calculate the magnetic moment
via
\begin{eqnarray}
\vec{\mu} &=& {1 \over 2} \int{d^3 x  (\vec{x} \times \vec{j})} \\
\nonumber
 &=& 2 q \kappa m_z \pi^2
\int_{0}^{\infty}{dr r F(r) G(r) } \hat{z}. \label{mm}
\end{eqnarray}
Note in Table \ref{table1} that we obtain the result $|\mu_z| \simeq {q
\over 2}$ when we choose $\kappa = 1$.

\section{Full Maxwell-Dirac Solution}
\label{sec5}

{}From our calculations in the previous section we expect that the inclusion
of the magnetic interaction will result in the emergence of a small
non-trivial angular dependence on the $\theta$ coordinate.  We also expect
the solution to maintain its axial symmetry and proceed in cylindrical
coordinates $(\rho,z,\phi)$, assuming a wavefunction of the form
\begin{equation}
 \psi = \left (\matrix{
{\psi_1}(\rho,z) {e}^{i (m_z-{1 \over 2}) \phi} \cr
{\psi_2}(\rho,z) {e}^{i (m_z+{1 \over 2}) \phi}  \cr
-i {\psi_3}(\rho,z) {e}^{i (m_z-{1 \over 2}) \phi} \cr
-i {\psi_4} (\rho,z) {e}^{i (m_z+{1 \over 2}) \phi} \cr
}\right ) \label{csw}
\end{equation}
and potentials of the form $\Phi= {1 \over q} V(\rho,z),\vec{A}= {1 \over
q} \it{A}(\rho,z)\hat{\phi}$.  Equations (\ref{pe}-\ref{j4}) now reduce to
\widetext
\begin{eqnarray}
 (\partial_\rho^2+{1 \over {4 \rho^2}} +\partial_z^2)(\sqrt{\rho} V) &=&
	 - {4 \pi \alpha \sqrt{\rho}}
({\psi_1}^{2}+{\psi_2}^{2}+{\psi_3}^{2}+{\psi_4}^{2}), \label{code} \\
 (\partial_\rho^2-{3 \over {4 \rho^2}} +\partial_z^2)(\sqrt{\rho} \it{A})
&=&
	 {8 \pi \alpha \sqrt{\rho}} ({\psi_1}{\psi_4}-{\psi_2}{\psi_3}),
\label{acode} \end{eqnarray}
and the Hamiltonian (\ref{vde}) becomes
\begin{equation}
\pmatrix{
V + 1 &  0 & -\partial_z & -\partial_\rho - {(m_z+{1 \over 2}) \over \rho}
+ \it{A} \cr
 0 & V + 1 & -\partial_\rho+ {(m_z-{1 \over 2}) \over \rho} - \it{A} &
\partial_z  \cr
\partial_z & \partial_\rho + {(m_z+{1 \over 2}) \over \rho}  - \it{A} & V
- 1 & 0  \cr
\partial_\rho- {(m_z-{1 \over 2}) \over \rho} + \it{A} & -\partial_z & 0 &
V - 1 \cr
}
. \label{cwode} \end{equation}
We symmetrize (\ref{cwode}) by the substitution $\Psi_\alpha \equiv
\sqrt{\rho} \psi_\alpha$, giving us \begin{equation}
E \left (\matrix{
{\Psi_1}\cr
{\Psi_2}  \cr
{\Psi_3} \cr
{\Psi_4} \cr}\right ) = \pmatrix{
V + 1 &  0 & -\partial_z & -\partial_\rho - {m_z \over { \rho}} + \it{A}
\cr
 0 & V + 1 & -\partial_\rho+ {m_z \over { \rho}} - \it{A} & \partial_z
\cr
\partial_z & \partial_\rho + {m_z \over { \rho}}  - \it{A} & V - 1 & 0
\cr
\partial_\rho- {m_z \over { \rho}} + \it{A} & -\partial_z & 0 & V - 1 \cr
}
\left (\matrix{
{\Psi_1}\cr
{\Psi_2}  \cr
{\Psi_3} \cr
{\Psi_4} \cr}\right )
. \label{scwode} \end{equation}
\narrowtext
Note that our wavefunctions in both cases\ (\ref{swave}) and\ (\ref{csw})
are eigenfunctions of the $z$ component of the angular momentum $J_z$ with
eigenvalue $m_z = \pm {1 \over 2}$, where
\begin{equation} J_z \equiv L_z + {1 \over 2} \Sigma_z, \label{jz}
\end{equation}
in which $\vec{L} \equiv -i(\vec{x} \times \vec{\nabla})$ and
\begin{equation} \vec{\Sigma} \equiv \pmatrix{
\vec{\sigma} & 0 \cr
0 & \vec{\sigma} \cr
}.  \label{sig}
\end{equation}
The full system\ (\ref{code}-\ref{scwode}) is once again spin degenerate
since the change to $m_z = -m_z'$ produces the same system with the change
($\it{A} = -\it{A'}$, ${\Psi_1} = -{\Psi_2'}$, ${\Psi_2} = {\Psi_1'}$,
${\Psi_3} =  {\Psi_4'}$, ${\Psi_4} = -{\Psi_3'}$).  Although our
scalar-Dirac wavefunctions\ (\ref{swave}) are eigenfunctions of  $K \equiv
\gamma^0 (\vec{\sigma} \cdot \vec{L}+1)$ with eigenvalue $\kappa$, our
solutions of the non-spherically symmetric case will vary slightly from
these eigenfunctions.

Figure~\ref{d2} shows a numerical solution to the full system for the same
parameter choice as in Fig.~\ref{fg}.  Note that the angular dependence is
virtually indistinguishable from the approximate solution and that the
energy agrees reasonably well with the first order perturbation
approximation.

\section{Conclusion}

We have found a class of solutions to the full Maxwell-Dirac equations and
good approximate solutions via a scalar-Dirac equation.  In practice, the
approximate solution is easier to work with and provides better accuracy
for most calculations.

The issue of stability has not been directly addressed.  The success of
our iterative solution method suggests that each solution is stable in
regard to slow collapse or expansion; however, we suspect that each
solution will be unstable via radiative transitions to states of large
negative energy, as is any bound state solution to the Dirac equation.

The interpretation of our solution is not immediately clear.  Several
authors have used similar solutions to construct hadrons from interacting
quarks \cite{bar,lee}; however, these solutions were to systems of
nonlinear scalar fields interacting with the Dirac field and the solutions
were essentially a result of the nonlinear scalar self-interaction rather
than the coupling term.  As it exists now, we see that our class of
solutions to the Maxwell-Dirac system may not be immediately interpreted
as representing the leptons.  This is clear from the large value of $<r>$
and negative value for $E$ for $\alpha={1 \over 137}$ as well as from the
experimental fact that weak interactions play the starring role in lepton
transitions.  It is not inconceivable that our solution could represent
the leptons if we rework our analysis to accommodate the weak interaction.
 Such an undertaking would present an interesting avenue for future work.

\acknowledgments

I would like to express my appreciation to Roger Dashen and Henry
Abarbanel for their encouragement and guidance and to the Hopgood
Foundation and the San Diego chapter of the ARCS Foundation for their
support.

\begin{figure}
\caption{A normalized, localized solution to our scalar-Dirac system for
the choice of $\alpha=2.7$, $\kappa = 1$, and $n=0$, showing $f(r)$,
$g(r)$, and ${\it V}(r)$ scaled down by a factor of $6$ to be fully
visible.  Note the asymptotic behavior of $\it{V}(r) \simeq {\alpha \over
r}$ for large $r$.  The expectation value $<r>$ is also shown.  All values
are in natural units.  The one dimensional mesh was discretized into $200$
points for this calculation.}
\label{fg}
\end{figure}

\begin{figure}
\caption{Energy levels as a function of $\alpha$ for our scalar-Dirac
approximation.  The dashed curves represent deviations due to the magnetic
interaction calculated via the first order perturbation.  All values shown
are in natural units.}
\label{e}
\end{figure}

\begin{figure}
\caption{A normalized, localized solution to the Max\-well\--Dir\-ac
system for the choice of $\alpha=2.7$, showing the four components of the
wavefunction ${\psi_i}(\rho,z)$, the potential ${\it V}(\rho,z)$, and the
$\hat{\phi}$ component of the vector potential ${\it A}(\rho,z)$.  The
contours go from min(light) to max(dark) and all values are in natural
units.  The two dimensional mesh was discretized into $60 \times
(2\times60+1)$ points for this calculation.  The energy was calculated as
$E_2=0.22$, to be compared with that obtained from the 1-D approximation,
$E+\Delta E_M=0.30$.}
\label{d2}
\end{figure}

\mediumtext
\begin{table}
\caption{Numerical results for our scalar-Dirac approximation on a $200$
point mesh for several choices of $\alpha$, $n$, and $\kappa$.  All values
shown are in natural units.  \label{table1}}
\begin{tabular}{ccccccc}
$\alpha$&$n$&$\kappa$&$E$&${\Delta E}_M$&$<r>$&$\mu_z /(q m_z)$\\
\tableline
1/137& 0& 1& $-1+9 \times 10^{-6}$& -0.00& 572& -1.2 \\
0.1& 0& 1& $-1+2 \times 10^{-3}$& -0.00& 42& -1.2 \\
1.0& 0& 1& -0.83& -0.00& 4.1& -1.1 \\
2.0& 0& 1& -0.28& -0.05& 1.9& -1.0 \\
2.4& 0& 1& 0.09& -0.10& 1.5& -0.9 \\
2.7\tablenotemark[1]& 0& 1& 0.46& -0.16& 1.2& -0.8 \\
3.0& 0& 1& 0.93& -0.24& 1.0& -0.7 \\
1/137& 1& -1& $-1+3 \times 10^{-6}$& -0.00& 1710& -0.4 \\
2.0& 1& -1& -0.75& -0.00& 5.4& -0.3 \\
3.4& 1& -1& 0.06& -0.06& 2.0& -0.2 \\
3.8& 1& -1& 0.89& -0.14& 1.2& -0.1 \\
3.0& 1& 1& -0.68& -0.00& 6.2& -1.1 \\
5.0& 1& 1& 0.32& -0.05& 2.4& -0.9 \\
5.5& 1& 1& 0.90& -0.10& 1.8& -0.8 \\
2.0& 2& -1& -0.93& -0.00& 18& -0.4 \\
5.8& 2& -1& 0.02& -0.03& 3.5& -0.2 \\
6.0& 2& -1& 0.18& -0.03& 3.1& -0.2 \\
6.0& 2& 1& -0.37& -0.01& 6.2& -1.0 \\
6.0& 3& 1& -0.71& -0.00& 13& -1.1 \\
\end{tabular}
\tablenotetext[1]{Solution shown in Fig.~\ref{fg}.}
\end{table}
\narrowtext
\end{document}